\documentclass[12pt,twoside,openbib]{article}

\usepackage{a4,amsmath,eucal,exscale}

\newcommand {\al}   {\alpha}
\newcommand {\bt}   {\beta}
\newcommand {\g }   {\gamma}
\newcommand {\G }   {\Gamma}
\newcommand {\dl}   {\delta}
\newcommand {\e }   {\epsilon}
\newcommand {\z }   {\zeta}
\newcommand {\et}   {\eta}

\newcommand {\lm}   {\lambda}
\newcommand {\m }   {\mu}

\newcommand {\h }   {\chi}

\newcommand {\om}   {\omega}

\newcommand {\Om}   {\Omega}

\newcommand {\pl}   {\partial}
\newcommand {\nb}   {\nabla}

\newcommand{\Conn}{\Omega}
\newcommand{\Tor}{t}
\newcommand{\tor}{\hat t}

\newcommand{\maps}{\text{:}\;}

\newcommand{\diag}{\operatorname{diag}}
\newcommand{\tr}{\operatorname{tr}}
\newcommand{\sdet}{\operatorname{sdet}}

\newcommand{\clmn}[2]{\left(
    \begin{array}{rr}
      #1 \\
      #2 
    \end{array}\right)
  }

\newcommand{\mtrx}[4]{\left(
    \begin{array}{cc}
      #1 & #2 \\
      #3 & #4
    \end{array}\right)
  }

\renewcommand{\^}{{}^}
\renewcommand{\_}{\!{}_}

\begin{document}

\begin{titlepage}
  \renewcommand{\thefootnote}{\fnsymbol{footnote}}

\begin{center}
  
  \hspace*{\fill} TUW-96-26

  \hspace*{\fill} \textbf{revised January 1998}
  
  \vspace*{\fill}

  \textbf{\Large Generalized Supergravity in Two Dimensions}
        
  \vspace{7ex}
        
  M.~F.~Ertl\footnotemark[1], M.~O.~Katanaev\footnotemark[2],
  W.~Kummer\footnotemark[3]
  
  \vspace{7ex}
  
  {\footnotemark[1]\footnotemark[3]\footnotesize Institut f\"ur
    Theoretische Physik \\ Technische Universit\"at Wien \\ Wiedner
    Hauptstr.  8--10, A-1040 Wien, Austria}
  
  \vspace{2ex}
  
  {\footnotemark[2]\footnotesize Steklov Mathematical Institute \\
    Gubkin St. 8, Moscow 117966, Russia}
             
  \footnotetext[1]{E-mail: \texttt{ertl@tph.tuwien.ac.at}}
  \footnotetext[2]{E-mail: \texttt{katanaev@mi.ras.ru}}
  \footnotetext[3]{E-mail: \texttt{wkummer@tph.tuwien.ac.at}}

\end{center}

\vspace{7ex}

\begin{abstract}
  Among the usual constraints of (1,1) supergravity in $d = 2$ the
  condition of vanishing bosonic torsion is dropped. Using the
  \emph{inverse} supervierbein and the superconnection considerably
  simplifies the formidable computational problems. It allows to solve
  the constraints for those fields \emph{before} taking into account
  the (identically fulfilled) Bianchi identities.  The relation of
  arbitrary functions in the seminal paper of Howe to supergravity
  multiplets is clarified. The local supersymmetry transformations
  remain the same, but, somewhat surprisingly, the transformations of
  zweibein and Rarita-Schwinger field decouple from those of the
  superconnection multiplet. A method emerges naturally, how to
  construct `non-Einsteinian' supergravity theories with nontrivial
  curvature and torsion in $d = 2$ which, apart from their intrinsic
  interest, may be relevant for models of super black holes and for
  novel generalizations in superstring theories. Several explicit
  examples of such models are presented, some of which immediately
  allow a dilatonic formulation for the bosonic part of the action.
\end{abstract}

\vspace*{\fill}\vspace*{\fill}

\renewcommand{\thefootnote}{\arabic{footnote}}
\setcounter{footnote}{0}
\end{titlepage}


\section{Introduction}
\label{sec:intro}

Despite the fact that no tangible direct evidence for supersymmetry so
far has been discovered in Nature, supersymmetry since its discovery
\cite{supersymmetry} managed to retain continual interest: within the
aim to arrive at a fundamental `theory of everything' first in
supergravity \cite{supergravity} in $d = 4$, then in generalizations
to higher dimensions of higher N \cite{vafa96}, and finally
incorporated as a low energy limit of superstrings \cite{witten95} or,
recently, of even more fundamental theories \cite{sezgin97} in 11
dimensions.

Even before the advent of strings and superstrings the importance of
studies in $1 + 1$ `space-time' had been emphasized \cite{howe79} in
connection with the study of possible superspace formulations
\cite{fayet77}. To the best of our knowledge to this day, however, no
attempt has been made to generalize the supergravity formulation of
(trivial) Einstein-gravity in $d = 2$ to the consideration of
two-dimensional $(1, 1)$ supermanifolds for which the condition of
vanishing (bosonic) torsion is removed. Only attempts to formulate
theories with higher power of curvature (at vanishing torsion) seem to
exist \cite{hindawi-ovrut-waldram96}. There seem to be only very few
exact solutions of supergravity in $d = 4$ as well
\cite{super-black-hole}.

Especially at times, when the number of arguments in favour of the
existence of an, as yet undiscovered, fundamental theory increase
\cite{vafa96} it may seem appropriate to also exploit --- if possible
--- \emph{all} (super-)geometrical generalizations of the
two-dimensional stringy world sheet. That such an undertaking can be
(and indeed is) successful is suggested by the recent much improved
insight, achieved for all (non-supersymmetric) two-dimensional
diffeomorphism invariant theories, including dilaton theory, and
permitting besides curvature also torsion \cite{dynamic-torsion} in
the most general manner
\cite{non-einsteinian-gravity,classical-and-quantum-gravity,general-dilaton-gravity}.
In the absence of matter-fields (non-geometrical degrees of freedom)
all these models are integrable at the classical level and admit the
analysis of all global solutions
\cite{katanaev-torsion,classical-and-quantum-gravity}. Integrability
of two-dimensional gravity coupled to chiral fermions was demonstrated
in \cite{kummer92,chiral-fermion-gravity}. Even the general aspects of
quantization of any such theory now seem to be well understood
\cite{kummer-schwarz92,non-einsteinian-gravity,haider-kummer94,kummer-liebl-vassilevich97}.
By contrast in the presence of matter, and if black holes like
singularities occur in such models, integrable solutions are known
only for very few cases. These include interactions with fermions of
one chirality \cite{kummer92}, and if scalar fields are present, only
the dilaton black hole \cite{dilaton-black-hole} and models which have
asymptotical Rindler behaviour \cite{fabbri-russo96}.  Therefore, a
supersymmetric extension of the matterless case suggests that the
solvability may carry over, in general. Then `matter' could be
represented by superpartners of the geometric bosonic field variables.

A straightforward approach would consist in a repetition of the
calculation of \cite{howe79} with nonvanishing bosonic torsion.
However, already for the simpler case treated in that reference, the
computational problems are considerable. We found it by far more
suitable to solve the constraints \emph{first}.  Using the
\emph{inverse} supervierbein and superconnection --- with conventional
gauge-fixing --- this task reduces to an algebraic (albeit still
lengthy) problem. By construction this solution fulfils the
Bianchi identities, but the latter are used nevertheless to determine
the components of supertorsion and supercurvature.

Section \ref{sec:supergeometry} is devoted to the general definitions
of superspace used in our present paper. In section
\ref{sec:gauge-fixing} after introducing the torsion constraints
\emph{without} the requirement of vanishing bosonic torsion we fix the
gauge in a way which later will turn out to be the correct one so that
the remaining supergravity transformations are indeed the local
generalizations of rigid supersymmetry.  The constraints are solved in
section \ref{sec:constraints} yielding the supermultiplets of vierbein
and superconnection in terms of an arbitrary supergravity and
a superconnection (or supertorsion) multiplet.  In section
\ref{sec:supercurvature} we compute the torsion and curvature
components of superspace.  The residual symmetry transformations of
the supermultiplets of vierbein and Lorentz connection are contained
in section \ref{sec:symmetries}, whereas finally in section
\ref{sec:lagrangians} several simple examples of general
(non-Einsteinian) supergravity actions within the superspace approach
are constructed.  In two appendices we describe notations and
conventions.


\section{Geometry of superspace}
\label{sec:supergeometry}

In $d = 2$ we consider a superspace with two commuting (bosonic) and
two anticommuting (Grassmann or spinor) coordinates
$z^M=\{x^m,\theta^\m\}$ where lower case Latin ($ m=0,1$) and Greek
indices ($\m=1,2$) denote commuting and anticommuting coordinates,
respectively:
\begin{equation}
  z^Mz^N=z^Nz^M(-1)^{MN}.
\end{equation}
Within our conventions for Majorana spinors (cf.\ appendix
\ref{app:spinors}) the first anticommuting element of the Grassmann
algebra is supposed to be real, $\theta^{1*}=\theta^1$, while the
second one is purely imaginary, $\theta^{2*}=-\theta^2$.

Our construction is based on differential geometry of superspace. We
shall not deal with subtle mathematical definitions
\cite{supergeometry}.  For our purpose it is sufficient to
follow one simple working rule allowing to generalize ordinary
formulas of differential geometry to superspace.

The exterior derivative operator in superspace
\begin{equation}
  d=dz^M\pl_M
\end{equation}
is invariant under arbitrary nondegenerate coordinate changes
$z^M\rightarrow z'\^{M}(z)$:
\begin{equation}
  \label{ecooch}
  dz^M \pl_M = dz'\^{M} \frac{\pl z^L}{\pl z'\^{M}} \frac{\pl
    z'\^{N}}{\pl z^L} \pl_{N'}
\end{equation}
Summation over repeated indices is assumed, and derivatives are always
supposed to be the left derivatives. From (\ref{ecooch}) follows our
simple basic rule: Any formula of differential geometry in ordinary
space can be taken over to superspace if the summation is always
performed from the upper left corner to the lower right one with no
indices in between (`ten to four'), and the order of the indices in
each term of the expression must be the same. Otherwise an appropriate
factor $(-1)$ must be included. E.\,g.\ the invariant interval reads
\begin{equation}
  ds^2=dz^M\! \otimes dz^N G_{NM} = dz^M\! \otimes dz^N G_{MN}
  (-1)^{MN},
\end{equation}
where $G_{MN}$ is the superspace metric. This metric can be used to
lower indices of a vector field,
\begin{equation}
  \label{elowin}
  V_M=V^NG_{NM}=G_{MN}V^N(-1)^N.
\end{equation}
The generalization to an arbitrary tensor is obvious.  Defining the
inverse metric according to the rule
\begin{equation}
  V^M=G^{MN}V_N=V_NG^{NM}(-1)^N,
\end{equation}
and demanding that sequential lowering and raising indices to be the
identical operation yields the main property of the inverse metric
\begin{equation}
  G^{MN} G_{NP} = \dl_P\^M (-1)^{MP} = \dl_P\^M (-1)^M = \dl_P\^M
  (-1)^P.
\end{equation}
The last identities follow from the diagonality of the Kronecker
symbol $\dl_P\^M = \dl_P^M$. Thus the inverse metric is not an inverse
matrix in the usual sense.  From (\ref{elowin}) the quantity
\begin{equation}
  V^2=V^MV_M=V_MV^M(-1)^M
\end{equation}
is a scalar, (but e.\,g.\ $V_M V^M$ is not!).

We assume that our superspace is equipped with a Riemann-Cartan
geometry that is with a metric and with a metrical connection
$\G_{MN}\^P$. The latter defines the covariant derivative of a tensor
field. Covariant derivatives of a vector $V^N$ and covector $V_N$ read
as
\begin{align}
  \nb_MV^N &= \pl_MV^N+V^P\G_{MP}\^N(-1)^{PM}, \\
  \nb_MV_N &= \pl_MV_N-\G_{MN}\^PV_P.
\end{align}
The metricity condition for the metric is
\begin{equation}                                          
\label{emecom}
\nb_MG_{NP}=\pl_MG_{NP}-\G_{MN}\^RG_{RP}-\G_{MP}\^RG_{RN}(-1)^{NP}=0.
\end{equation}
The action of an (anti)commutator of covariant derivatives,
\begin{equation}
  [\nb_M,\nb_N\} = \nb_M\nb_N-\nb_N\nb_M(-1)^{MN}
\end{equation}
on a vector field (\ref{elowin}),
\begin{equation}
  [\nb_M,\nb_N\}V_P=-R_{MNP}\^RV_R-T_{MN}\^R\nb_RV_P,
\end{equation}
is defined in terms of curvature and torsion:
\begin{align}                                        
  R_{MNP}\^R &= \pl_M \G_{NP}\^R - \G_{MP}\^S \G_{NS}\^R
  (-1)^{N(S+P)} - (M \leftrightarrow N) (-1)^{MN}, \label{ecurvs} \\
  T_{MN}\^R &= \G_{MN}\^R - \G_{NM}\^R (-1)^{MN} \label{etorss}
\end{align}

In our construction we use Cartan variables: the superspace vierbein
$E_M\^A$ and the superconnection $\Conn_{MA}\^B$.  Capital Latin
indices from the beginning of the alphabet $(A={a,\al})$ transform
under the Lorentz group as a vector $(a=0,1)$ and spinor $(\al=1,2)$,
respectively. Cartan variables are defined by
\begin{equation}
  \label{eviers}
  G_{MN}=E_M\^AE_N\^B\et_{BA}(-1)^{AN},
\end{equation}
and the metricity condition
\begin{equation}
 \label{elocos}
 \nb_ME_N\^A=\pl_ME_N\^A-\G_{MN}\^PE_P\^A
 +E_N\^B\Conn_{MB}\^A(-1)^{M(B+N)}=0.
\end{equation}
Raising and lowering of the anholonomic indices $(A,B,\dots)$ is
performed by the superspace Minkowski metric
\begin{equation}
  \label{eminms}
  \et_{AB} = \mtrx{\et_{ab}}{0}{0}{\e_{\al\bt}}, \quad
  \et^{AB} = \mtrx{\et^{ab}}{0}{0}{\e^{\al\bt}},
\end{equation}
consisting of the two-dimensional Minkowskian metric
$\et_{ab}=\et^{ab}=\diag(+-)$ and $\e_{\al\bt}$, the totally
antisymmetric (Levi-Civita) tensor defined in appendix
\ref{app:spinors}. The Minkowski metric and its inverse (\ref{eminms})
in superspace obey
\begin{equation}
  \et_{AB}=\et_{BA}(-1)^A, \quad \et^{AB}\et_{BC}=\dl_C\^A(-1)^A.
\end{equation}
The transformation of anholonomic indices $(A,B,\dots)$ into holonomic
indices $(M,N,\dots)$ and vice versa is performed using the
supervierbein and its inverse $E_A\^M$ defined as
\begin{equation}
  \label{einvsv}
  E_A\^ME_M\^B=\dl_A\^B, \quad E_M\^AE_A\^N=\dl_M\^N.
\end{equation}
The metric (\ref{eminms}) is invariant under the Lorentz group acting
on tensor indices from the beginning of the alphabet. In fact
(\ref{eminms}) is not unique in this respect because $\e_{\al\bt}$ may
be multiplied by an arbitrary nonzero factor. This may represent a
freedom to generalize our present approach. In fact, in order to have
a correct dimension of all terms in the line element of superspace,
that factor should carry the dimension of length. A specific choice
for it presents a freedom in approaches to supersymmetry. In the
following this factor will be suppressed. Therefore any apparent
differences in dimensions between terms below are not relevant.

The metricity condition (\ref{elocos}) formally establishes a
one-to-one correspondence between the metrical connection $\G_{MN}\^P$
and the superconnection $\Conn_{MA}\^B$.  Together with (\ref{emecom})
it implies
\begin{equation}
  \nabla_M \eta_{AB} = 0
\end{equation}
and the symmetry property
\begin{equation}
  \label{esupco}
  \Conn_{MAB} + \Conn_{MBA} (-1)^{AB} = 0.
\end{equation}
In general, the superconnection $\Conn_{MA}\^B$ is not related to
Lorentz transformations alone. The Lorentz connection in superspace,
as will be seen from section \ref{sec:gauge-fixing}, must have a
specific form and is defined (in $d = 2$) by $\Conn_M$, a superfield
with one vector index,
\begin{equation}
  \label{elorcs}
  \Conn_{MA}\^B = \Conn_M L_A\^B,
\end{equation}
where
\begin{equation}
  \label{eantis}
  L_A\^B = \mtrx{\e_a\^b}{0}{0}{-\frac12 \g^5\_\al\^\bt}
\end{equation}
contains the Lorentz generators in the bosonic and fermionic sectors.
Here the factor in front of $\g^5$ is fixed by the requirement that
under Lorentz transformations $\g$-matrices are invariant under
simultaneous rotations of vector and spinor indices. Definition and
properties of $\g$-matrices are given in appendix \ref{app:spinors}.
$L_A\^B$ has the properties
\begin{equation}
  L_{AB} = -L_{BA} (-1)^A, \quad L_A\^B L_B\^C =
  \mtrx{\delta_a\^c}{0}{0}{\frac{1}{4} \delta_{\alpha}\^{\gamma}},
  \quad \nabla_M L_A\^B = 0.
\end{equation}
The superconnection $\Conn_{MA}\^B$ in the form (\ref{elorcs}) is very
restricted, because the original 32 independent superfield components
for the Lorentz superconnection reduce to 4. As a consequence,
(\ref{elocos}) with (\ref{elorcs}) also entails restrictions on the
metric connection $\G_{MN}\^P$.

In terms of the connection (\ref{elorcs}) covariant derivatives of a
Lorentz supervector read
\begin{align}
  \nb_M V^A &= \pl_M V^A + \Conn_M V^B L_B\^A, \\
  \nb_M V_A &= \pl_M V_A - \Conn_M L_A\^B V_B.
\end{align}
The (anti)commutator of covariant derivatives,
\begin{equation}
  [\nb_M,\nb_N\}V_A =-R_{MNA}\^BV_B-T_{MN}\^P\nb_PV_A,
\end{equation}
is defined by the same expressions for curvature and torsion as given
by (\ref{ecurvs}) and (\ref{etorss}), which in Cartan variables become
\begin{align}
  R_{MNA}\^B &= \pl_M \Conn_{NA}\^B - \Conn_{MA}\^C \Conn_{NC}\^B
  (-1)^{N(A+C)} - (M \leftrightarrow N) (-1)^{MN}, \label{ecarcu} \\
  T_{MN}\^A &= \pl_M E_N\^A + E_N\^B \Conn_{MB}\^A (-1)^{M(B+N)} - (M
  \leftrightarrow N) (-1)^{MN}. \label{ecarto}
\end{align}
In terms of the Lorentz connection (\ref{elorcs}) the curvature does
not contain quadratic terms
\begin{equation}
  \label{ecurts}
  R_{MNA}\^B = \left(\pl_M \Conn_N - \pl_N \Conn_M (-1)^{MN}\right)
  L_A\^B = F_{MN} L_A\^B.
\end{equation}

In the calculations below we have found it extremely convenient to
work directly in the anholonomic basis
\begin{displaymath}
  D_A=E_A\^M\pl_M,
\end{displaymath}
defined by the inverse supervierbein. In this basis curvature
(\ref{ecurts}) and torsion (\ref{ecarto}) become
\begin{align}
  F_{AB} &= -C_{AB}\^C \Conn_C + D_A \Conn_B - D_B \Conn_A
  (-1)^{AB}, \label{etoran} \\
  T_{AB}\^C &= -C_{AB}\^C + \Conn_A L_B\^C - \Conn_B L_A\^C (-1)^{AB},
  \label{ecuran}
\end{align}
where
\begin{displaymath}
  C_{AB}\^C = \left(E_A\^N \pl_N E_B\^M - E_B\^N \pl_N E_A\^M
  (-1)^{AB}\right) E_M\^C
\end{displaymath}
are the coefficients of anholonomicity. Beside the inverse
supervierbein $E_A\^M$ also the use of the Lorentz superconnection in
the anholonomic basis, $\Conn_A = E_A\^M \Conn_M$ has proved to be
crucial for our approach.
  
Ricci tensor and scalar curvature of the manifold are
\begin{align}
  R_{AB} &= R_{ACB}\^C (-1)^{C(B+C)} = -L_B\^C F_{CA} (-1)^{AB},
  \label{esrict} \\
  R &= R_A\^A (-1)^A = -L^{AB} F_{BA}. \label{esuscu}
\end{align}
Curvature and torsion satisfy the Bianchi identities
\begin{multline}
  \label{ebiacs}
  \nb_A R_{BCDE} + \nb_B R_{CADE} (-1)^{A(B+C)} + \nb_C R_{ABDE}
  (-1)^{C(A+B)} \\
  + T_{AB}\^F R_{FCDE} + T_{BC}\^F R_{FADE} (-1)^{A(B+C)} + T_{CA}\^F
  R_{FBDE} (-1)^{C(A+B)} = 0
\end{multline}
and
\begin{multline}
  \label{ebiats}
  \nb_A T_{BCD} + \nb_B T_{CAD} (-1)^{A(B+C)} + \nb_C T_{ABD}
  (-1)^{C(A+B)} = \\
  R_{ABCD} + R_{BCAD} (-1)^{A(B+C)} + R_{CABD} (-1)^{C(A+B)} \\
  -T_{AB}\^E T_{ECD} - T_{BC}\^E T_{EAD} (-1)^{A(B+C)} - T_{CA}\^E
  T_{EBD} (-1)^{C(A+B)}
\end{multline}
where with $R_{ABC}\^D = F_{AB} L_C\^D$ the Bianchi identity
(\ref{ebiacs}) simplifies considerably
\begin{multline}
  \label{ebiass}
  \nb_A F_{BC} + \nb_B F_{CA} (-1)^{A(B+C)} + \nb_C F_{AB}
  (-1)^{C(A+B)} \\
  + T_{AB}\^D F_{DC} + T_{BC}\^D F_{DA} (-1)^{A(B+C)} + T_{CA}\^D
  F_{DB} (-1)^{C(A+B)} = 0.
\end{multline}


\section{Supergravity Constraints and Gauge Fixing}
\label{sec:gauge-fixing}

Generally superdiffeomorphism invariant theories may be constructed
easily from supervierbein and superconnection.  However their flat
limit does not show rigid supersymmetry.  What is commonly defined to
be `supergravity' is the restricted case which is invariant under
diffeomorphisms in the bosonic subspace, local Lorentz boosts, and
local supersymmetry transformations mixing bosonic and fermionic
fields, all parameters depending only on the \emph{bosonic}
coordinates ($x^m$).  As a consequence of the uniqueness theorem of
suitably redefined supersymmetry transformations in rigid superspace
\cite{haag-lopuszanski-sohnius75} it is natural to postulate
\begin{equation}                                       \label{etorco}
  T_{\al\bt}\^a=2i(\g^a\e)_{\al\bt}, \quad T_{\al\bt}\^\g=0.
\end{equation}

These constraints are identical to the first two standard constraints,
also imposed by Howe \cite{howe79}. We find, though, that the third
constraint $T_{ab}\^c=0$ \cite{howe79} specifying the bosonic part of
the torsion to be `Einsteinian' is not mandatory for supersymmetry
transformations and may be dropped. This leads to a special type of
Riemann-Cartan superspace, where we retain the `maximal amount' of
(1,1)-superspace structure in the tangent space.  In particular,
(\ref{etorco}) should be invariant under superdiffeomorphism and
Lorentz transformation. The first requirement is trivially fulfilled.
The second one produces precisely the restriction on $\Conn_{MA}\^B$
anticipated in section \ref{sec:supergeometry}.  The superpartner of
an inverse zweibein $e_a\^m$ is supposed to be a Rarita-Schwinger
field $\chi_a\^\mu$ carrying one vector and one spinor index. Within
the superspace approach to supergravity diffeomorphisms of the bosonic
sector and supersymmetry transformations must appear as subgroups of
the full superspace diffeomorphisms
\begin{equation}
  \label{essdif}
  z'\^M = w^M(z) = w^{(0)}\^M + \theta^\nu w^{(1)}\_\nu\^M + \frac12
  \theta\bar\theta w^{(2)}\^M,
\end{equation}
where $w^{(0,1,2)}$ are functions of $x^m$ only.  Since all symmetry
parameters of a proper supergravity model are assumed to depend on
$x^m$ alone it is natural to use the functions $w^{(1)}$ and $w^{(2)}$
to fix certain components of the supervierbein.  Under the
transformation (\ref{essdif}) the components $E_\al\^M$ transform as
\begin{equation}
  \begin{aligned}
    E'\_\al\^m &= E_\al\^n \pl_n w^m + E_\al\^\nu
    \left(w^{(1)}\_\nu\^m + \theta_\nu w^{(2)}\^m\right), \\
    E'\_\al\^\mu &= E_\al\^n \pl_n w^\mu + E_\al\^\nu
    \left(w^{(1)}\_\nu\^\mu + \theta_\nu w^{(2)}\^\mu\right).
  \end{aligned}
\end{equation}
Expanding the inverse supervierbein in $\theta$
\begin{displaymath}
  E_A\^M = E^{(0)}\_A\^M + \theta^\nu E^{(1)}\_{\nu A}\^M + \frac12
  \theta\bar\theta E^{(2)}\_A\^M,
\end{displaymath}
one easily sees that if $\det E^{(0)}\_\al\^\mu \ne 0$ then one may
always use the functions $w^{(1)}\_\nu\^M$ to fix
$E^{(0)}\_\alpha\^M$:
\begin{equation}
  \label{egacof}
  E^{(0)}\_\al\^m = 0, \quad E^{(0)}\_\al\^\mu = \dl_\al\^\mu.
\end{equation}
In the next step the functions $w^{(2)}\^m$ and $w^{(2)}\^\mu$ may be
used to get rid of the antisymmetric parts of the first order
components
\begin{equation}
  \label{egacos}
  E^{(1)}\_{\nu\al}\^m = E^{(1)}\_{\al\nu}\^m, \quad
  E^{(1)}\_{\nu\al}\^\mu = E^{(1)}\_{\al\nu}\^\mu.
\end{equation}

Under the Lorentz boost the superconnection transforms according to
\begin{equation}
  \label{labl8}
  \Conn'\_A = S^{-1}\_A\^B (\Conn_B - D_B W),
\end{equation}
where $S^{-1}\_A\^B$ is the inverse Lorentz transformation matrix
corresponding to the boost parameter
\begin{equation}
  \label{labl6}
  W = W^{(0)} + \theta^\nu W^{(1)}\_\nu + \frac12 \theta\bar\theta
  W^{(2)},
\end{equation}
which also represents a superfield. The freedom in choosing $W^{(1)}$
and $W^{(2)}$ allows us to set
\begin{equation}
  \label{egacot}
  \Conn^{(0)}\_\al = 0, \quad \Conn^{(1)}\_{\nu\al} =
  \Conn^{(1)}\_{\al\nu},
\end{equation}
respectively. This exhausts all group parameters to first and second
order in $\theta$. So the remaining free parameters of the symmetry
transformations are $w^{(0)}\^m$, $w^{(0)}\^\mu$, and $W^{(0)}$ which
are functions of $x^m$ alone and (as we shall see later) indeed
describe bosonic diffeomorphisms, local supersymmetry transformations,
and local Lorentz boosts, respectively. The expansion in $\theta$ of
the supervierbein and the Lorentz superconnection, including the gauge
fixing (\ref{egacof}), (\ref{egacos}) and (\ref{egacot}), may be
written as
\begin{alignat}{2}
  E_a\^m &= e_a\^m+\theta^\nu f_{\nu
    a}\^m+\frac12\theta\bar\theta g_a\^m, \label{labl2a} \\
  E_a\^\mu &= \chi_a\^\mu+\theta^\nu f_{\nu a}\^\mu
  +\frac12\theta\bar\theta g_a\^\mu, \\
  E_\al\^m &= \theta^\nu f_{\nu\al}\^m+\frac12\theta\bar\theta
  g_\al\^m, &\quad f_{\nu\al}\^m &= f_{\al\nu}\^m, \\
  E_\al\^\mu &= \dl_\al\^\mu+\theta^\nu f_{\nu\al}\^\mu
  +\frac12\theta\bar\theta g_\al\^\mu, &\quad
  f_{\nu\al}\^\mu &= f_{\al\nu}\^\mu, \label{labl2b} \\
  \Conn_a &= \om_a + \theta^\nu \bar u_{\nu a} + \frac12
  \theta\bar\theta \upsilon_a, \label{labl3a} \\
  \Conn_\al &= \theta^\nu \rho_{\nu\al} + \frac12 \theta\bar\theta
  \upsilon_\al, &\quad \rho_{\nu\al} &= \rho_{\al\nu}. \label{labl3b}
\end{alignat}
Here $e_a\^m$, $\chi_a\^\mu$, and $\om_a$ are the inverse zweibein,
the Rarita-Schwinger field and the (bosonic) Lorentz connection,
respectively.  Part of the other components will be found in the next
section as the solution to the torsion constraints while the rest are
needed to form supermultiplets for $e$, $\chi$, and $\om$.

The Wess-Zumino type gauge (\ref{egacof}), (\ref{egacos}) and
(\ref{egacot}) is similar to that used by Howe [6] but it is applied
to the \emph{inverse} supervierbein. After solution of the constraints
and inverting the supervierbein we shall get a result consistent with
his gauge and vice versa. So the gauges are equivalent. In fact, what
amounts to `fixing the gauge' is similar to what happens in the coset
approach to `gauging' supersymmetry into supergravity
\cite{nieuwenhuizen81}.


\section{Solution of the constraints}
\label{sec:constraints}

Writing the constraints (\ref{etorco}) in equivalent form
\begin{equation}
  T_{\al\bt}\^M=2i(\g^a\e)_{\al\bt}E_a\^M,
\end{equation}
or, after contracting with suitable $\gamma$-matrices (cf.\ appendix
\ref{app:spinors}),
\begin{align}
  T_{\al\bt}\^M (\e\g^5)^{\bt\al} &= 0, \label{etoco1} \\
  T_{\al\bt}\^M (\e\g_a)^{\bt\al} &= -4i E_a\^M, \label{etoco2}
\end{align}
one observes that (\ref{etoco2}) simply becomes the relation
expressing $E_a\^M$ in terms of what has been obtained by the solution
of (\ref{etoco1}). Thus only the latter relation must be solved, which
contains through $T_{\al\bt}\^M$ the components $E_\alpha\^M$ of the
inverse supervierbein and $\Conn_\alpha$ of the Lorentz
superconnection. As we shall see all components of
(\ref{labl2a})--(\ref{labl2b}) and (\ref{labl3b}) can be solved in
terms of the supermultiplet $e_a\^m$, $\chi_a\^\m$ and an arbitrary
scalar field $A$ without using the Bianchi identities at this point.
If instead the supervierbein $E_M\^A$ is considered as an independent
variable then the constraints (\ref{etorco}) either both contain the
vierbein and its inverse thus providing a major computational problem,
or, when written for $T_{MN}\^A$ require the knowledge of other
torsion components. In that last case, which had been exploited in
[6], the prior solution of the Bianchi identities is inevitable.

To solve equations (\ref{etoco1}), (\ref{etoco2}) we decompose
$\rho$ and $f$ in (\ref{labl2a})--(\ref{labl3b}) in the basis for
symmetric matrices:
\begin{equation}
  \label{labl4}
  \begin{aligned}
    \rho_{\nu\al} &= A (\g^5\e)_{\nu\al} + i \rho_a (\g^a\e)_{\nu\al},
    \\
    f_{\nu\al}\^m &= f^m (\g^5 \e)_{\nu\al} + f_a\^m (\g^a
    \e)_{\nu\al}, \\
    f_{\nu\al}\^\mu &= f^\mu(\g^5 \e)_{\nu\al} + f_a\^\mu (\g^a
    \e)_{\nu\al}.
  \end{aligned}
\end{equation}
In addition we separate (cf.\ appendix \ref{app:spinors}) the
vector-spinor according to
\begin{equation}
  \chi^a = \chi\gamma^a + \lambda^a, \quad \gamma^a\lambda_a = 0
\end{equation}
into a Rarita-Schwinger field $\lambda^a$ and a spinor $\chi$.  Then
to zeroth order in $\theta$ the equations (\ref{etoco1}) and
(\ref{etoco2}) yield
\begin{alignat}{2}
  \label{esozor}
  f^m &= 0, &\quad f_a\^m &= -i e_a\^m, \\
  f^\mu &= 0, &\quad f_a\^\mu &= -i \chi_a\^\mu.
\end{alignat}
To first order in $\theta$ one obtains
\begin{equation}
  \label{esofon}
  g_\al\^m = 2\bar\lm^m\_\al, \quad f_{\nu a}\^m = i (\g^m
  \bar\chi_a)_\nu,
\end{equation}
and
\begin{align}
  g_\al\^\mu &= 2\bar\lm^b\_\al \chi_b\^\mu - \frac12 A \dl_\al\^\mu +
  \frac{i}{2} \rho_b (\g^b\g^5)_\al\^\mu, \label{esoftw}
  \\
  f_{\nu a}\^\mu &= i (\g^b \bar\chi_a)_\nu \chi_b\^\mu - \frac{i}{2}
  A \g_{a\nu}\^\mu - \frac12 \rho_a \g^5\_\nu\^\mu. \label{esofth}
\end{align}
The transformation of the lower case Latin indices from the beginning
of the alphabet to the ones from the middle is performed by using the
zweibein, e.\,g.\ $\chi_m\^\al=e_m\^a\chi_a\^\al$, $\g_m=e_m\^a\g_a$.
The two-dimensional bosonic metric $g_{mn}=e_m\^ae_{na}$ is used for
raising and lowering the indices.

Eqs.\ (\ref{etoco1}), (\ref{etoco2}) in second order of $\theta$
determine two functions in the Lorentz superconnection
\begin{align}
  \rho_a &= -\e_a\^b c_b - 4i(\lm_a \g^5 \bar\chi),
    \label{efurho} \\
    \upsilon_\al &= 4\e^{mn}(\widetilde\nb_m\bar\chi_n)_\al+2c_b
    (\g^5\bar\lm^b)_\al -2i A (\g^5\bar\chi)_\al \label{efupsi}
\end{align}
where the trace of the anholonomicity coefficients $c_b$ and the
covariant derivative $\widetilde\nb_m$ are defined in appendix
\ref{app:anhol-basis}.

The remaining components of the inverse supervierbein
(\ref{labl2a})--(\ref{labl2b}) are also obtained from that order as
\begin{align}
  g_a\^m &= -2(\lm^m \bar\chi_a), \label{esorbb} \\
  g_a\^\mu &= i\e^{mn}(\widetilde\nb_m\chi_n\g_a\g^5)^\mu
  +\frac{i}{2}c_b(\chi_a\g^b)^\mu+A(\chi\g_a)^\mu. \label{esorbf}
\end{align}

Thus after some rearrangements the general solution to the constraints
(\ref{etorco}) in the gauge (\ref{egacof}), (\ref{egacos}) and
(\ref{egacot}) can be summarized as ($\lm^2 = \lm^a \bar\lm_a$)
\begin{align}
  E_a\^m &= e_a\^m+i(\theta\g^m\bar\chi_a)
  -\theta\bar\theta(\lm^m\bar\chi_a), \label{einsvf} \\
  E_a\^\mu &= \chi_a\^\mu+i(\theta\g^b\bar\chi_a)\chi_b\^\mu -\frac
  i2A(\theta\g_a)^\mu-\frac12\rho_a(\theta\g^5)^\mu \nonumber \\
  &\phantom{= M} + \frac12 \theta\bar\theta \left[i\e^{mn}
    (\widetilde\nb_m \chi_n \g_a \g^5)^\mu +\frac i2c_b (\chi_a
    \g^b)^\mu+A (\chi\g_a)^\mu\right], \label{einsvs} \\
  E_\al\^m &= i (\g^m \bar\theta)_\al + \theta\bar\theta
  \bar\lm^m\_\al, \label{einsvt} \\
  E_\al\^\mu &= \dl_\al\^\mu + i(\g^b \bar\theta)_\al \chi_b\^\mu
  -\frac12 \theta\bar\theta \left[\lm^2 \dl_\al\^\mu + \frac12 A
    \dl_\al\^\mu +\frac i2c_b \g^b\_\al\^\mu\right], \label{einsvg} \\
  \intertext{for the inverse supervierbein and} \Conn_a &= \om_a +
  (\theta \bar u_a) + \frac12 \theta\bar\theta
  \upsilon_a \label{esulob} \\
  \Conn_\al &= A (\theta \g^5 \e)_\al + i\rho_b (\theta \g^b \e)_\al
  \nonumber \\
  &\phantom{= M} + \frac12 \theta\bar\theta \left[ 4 \e^{mn}
    (\widetilde\nb_m \bar\chi_n)_\al + 2 c_b (\g^5 \bar\lm^b)_\al - 2i
    A(\g^5 \bar\chi)_\al\right], \label{esulof}
\end{align}
for the superconnection where $\rho_a$ by (\ref{efurho}) is a function
of $e_a\^m$ and $\chi_a\^\mu$.  The very existence of this solution
proves that the torsion constraints are indeed consistent with the
Bianchi identities, without having to solve the latter at all.
Inspecting (\ref{einsvf})--(\ref{esulof}) shows that after solution of
the constraints we are left with two `supermultiplets': the
supergravity multiplet ${\cal E} = \{ e_a\^m,\chi_a\^\mu,A \}$ and the
Lorentz connection supermultiplet $\Om_a = \{ \om_a, u_a\^\nu,
\upsilon_a \}$. The supergravity multiplet consists of the inverse
zweibein $e_a\^m$, the vector-spinor field $\chi_a\^\mu$, and the
scalar field $A$. The components of $\cal E$ originate from components
of different superfields whereas the Lorentz connection supermultiplet
represents one superfield $\Conn_a$. It does not enter the constraint
equation at all and so remains completely arbitrary. In the next
section we shall see that the parameters of local symmetry
transformations depend only on $\cal E$.

Comparing our result with the one by Howe \cite{howe79} we observe
that our function $A$ (up to a factor) is identical to $A$ of that
paper.  There it appeared as the first component of a scalar
superfield entering a general solution of the Bianchi identities.  Our
use of the inverse supervierbein as a primary field thus not only
allows to postpone the discussion of the Bianchi identities (thus
simplifying the calculations), but also clarifies the geometrical
meaning of $A$ as the scalar component of $\cal E$.


\section{Superspace curvature and torsion}
\label{sec:supercurvature}

In order to compute superspace curvature and torsion in terms of the
supermultiplets ${\cal E}$ and $\Omega$ one needs the explicit form
of the supervierbein $E_M\^A$. Solving one of the eqs.\ (\ref{einvsv})
we find the components
\begin{align}
  E_m\^a &= e_m\^a - 2i (\theta \g^a \bar\chi_m) + \frac12
  \theta\bar\theta A e_m\^a, \label{esuvbb} \\
  E_m\^\al &= - \chi_m\^\al + \frac{i}{2} A (\theta \g_m)^\al +
  \frac12 \rho_m (\theta \g^5)^\al \nonumber \\
  &\phantom{=} + \frac12 \theta\bar\theta \left[ -i \e^{np}
    (\widetilde\nb_n \chi_p \g_m \g^5)^\al + \lm^2 (\chi \g_m)^\al +
    \frac12 A (\chi\g_m)^\al + \frac32 A \lm_m\^\al \right],
  \label{esuvbf} \\
  E_\mu\^a &= i (\theta \g^a\e)_\mu, \label{esuvfb} \\
  E_\mu\^\al &= \dl_\mu\^\al \left( 1 - \frac14 \theta\bar\theta A
  \right). \label{esuvff}
\end{align}
Eqs.\ (\ref{esuvbb})--(\ref{esuvff}) in terms of components indeed
coincide (up to misprints) with those of Howe \cite{howe79}.

Now we are able to compute the anholonomic components of curvature and
torsion defined in section \ref{sec:supergeometry}.  Only at this
point it becomes useful to take the Bianchi identities into account.
For the constraints (\ref{etorco}) a straightforward, but tedious
calculation yields expressions for torsion
\begin{align}
  T_{\al\bt}\^\g &= 0, \label{eto1} \\
  T_{\al\bt}\^a &= 2i (\g^a \e)_{\al\bt}, \label{eto2} \\
  T_{\al a}\^\bt &= \frac{i}{2} S \g_{a\al}\^\bt + \frac12 \e_a\^b T_b
  \g^5\_\al\^\bt, \label{eto3} \\
  T_{\al a}\^b &= -T_{a\al}\^b = 0, \label{eto4} \\
  T_{ab}\^\al &= \frac12 \e_{ab} (\e \g^5)^{\al\bt} \nb_\bt S,
  \label{eto5} \\ 
  T_{ab}\^c &= \dl_a\^c T_b - \dl_b\^c T_a \label{eto6} \\
  \intertext{and curvature}
  F_{\al\bt} &= 2 S (\g^5 \e)_{\al\bt} + 2i T_a (\g^a \g^5
  \e)_{\al\bt}, \label{ecu1} \\ 
  F_{\al a} &= -F_{a\al} = i (\g_a \g^5)_\al\^\bt \nb_\bt S
  + \e_a\^b \nb_\al T_b, \label{ecu2} \\
  F_{ab} &= \e_{ab} \left[ -\frac12 \e^{\al\bt} \nb_\bt \nb_\al S +
  S^2 - \nb_c T^c + T_c T^c \right] \label{ecu3}
\end{align}
in terms of $S$ and $T_a$, a scalar and a vector superfield.  In order
to find out how they depend on the components of $E_M\^A$, eqs.\ 
(\ref{esuvbb})--(\ref{esuvff}), it suffices to compute the l.\,h.\,s.\ 
of (\ref{eto3}) or (\ref{ecu1}) and to compare both sides of the
equation.  The scalar superfield turns out to depend on $\cal E$ alone
\begin{multline}
  \label{escsuf}
  S = A + 2 \e^{mn} (\theta \g^5 \widetilde\nb_m \bar\chi_n) + 2i
  (\theta \bar\chi) \lm^2 - 2i A(\theta \bar\chi) \\
  - \frac12 \theta\bar\theta \left[ \frac12 \widetilde R - 4i \e^{mn}
    (\chi \g^5 \widetilde\nb_m \bar\chi_n ) + 4i \widetilde
    \nb_a(\lm^a \bar\chi) + 4 \chi^2 \lm^2 +A (\chi^a \bar\chi_a) +
    A^2 \right]
\end{multline}
where $\widetilde R$ is the curvature scalar for vanishing (bosonic)
torsion (cf.\ appendix \ref{app:anhol-basis}).  The vector superfield
is nothing else than the trace of the supertorsion
\begin{equation}
  T_B = T_{AB}\^A (-1)^{A+AB} = \left( T_b, T_\bt = 0 \right).
\end{equation}
The components of $T_a = -C_a + \e_a\^b \Conn_b$, where $C_a =
C_{ba}\^b$,
\begin{equation}
  \label{esutor}
  T_a = \Tor_a + (\theta\bar \tau_a) + \frac12 \theta\bar\theta s_a,
\end{equation}
depend on both $\cal E$ and $\Omega$,
\begin{align}
  \Tor_a &= \tor_a - 4i (\lm_a \bar\chi) \label{estocz} \\
  (\theta \bar\tau_a) &= 2i \e^{mn} (\theta \g_a \g^5 \widetilde\nb_m
  \bar\chi_n) - i \e_a\^b c_c (\theta \g^c \g^5 \bar\chi_b) \nonumber
  \\
  &\phantom{= M} + 4 \chi^2 (\theta \bar\lm_a) - A (\theta \bar\chi_a)
  + \e_a\^b (\theta \bar u_b), \label{estocf} \\
  s_a &= -\pl_a A + 4 \e^{mn} (\widetilde\nb_m \chi_n \g_a \g^5
  \bar\chi) - 2 \e_a\^b c_c (\lm^c \g^5 \bar\chi_b) \nonumber \\
  &\phantom{= M} + 2i A (\chi \bar\lm_a) + \e_a\^b \upsilon_b.
  \label{estocs}
\end{align}
Here $\tor_a$ is the trace of the torsion of the two-dimensional
bosonic subspace (cf.\ appendix \ref{app:anhol-basis}).  The
expressions for supertorsion and supercurvature obtained here will be
used in section \ref{sec:lagrangians} for the construction of
generalized supergravity Lagrangians.


\section{Symmetry transformations}
\label{sec:symmetries}

Since the gauge was not fixed completely several parameters are still
free: the zeroth components of superspace diffeomorphisms $w^{(0)M}$
(cf.\ (\ref{essdif})) and the zeroth component of the Lorentz rotation
$W^{(0)}$ in (\ref{labl6}). They determine the remaining symmetry
transformations. If our gauge fixing in section \ref{sec:gauge-fixing}
was a suitable one we should obtain in this way the correct local
transformations of supergravity.

Under infinitesimal superdiffeomorphisms parametrized by a vector
superfield $\xi^M(z)$, and by an infinitesimal Lorentz (super-)boost
with parameter $W(z)$ the inverse supervierbein and the anholonomic
components of the Lorentz superconnection obey the transformation
formulas
\begin{equation}
\label{labxx}
  \begin{aligned}
    \dl E_A\^M &= \xi^N \pl_NE_A\^M - E_A\^N \pl_N \xi^M - W L_A\^B
    E_B\^M, \\
    \dl \Conn_A &= \xi^N \pl_N \Conn_A - W L_A\^B \Conn_B - E_A\^M
    \pl_M W.
  \end{aligned}
\end{equation}

To find the explicit form of the remaining symmetry transformations
after the gauge fixing of section \ref{sec:gauge-fixing} we decompose
\begin{equation}
  \label{labl9}
  \begin{aligned}
    \xi^m &= \zeta^m + \theta^\nu k_\nu\^m + \frac12 \theta\bar\theta
    l^m, \\
    \xi^\mu &= \zeta^\mu + \theta^\nu k_\nu\^\mu + \frac12
    \theta\bar\theta l^\mu, \\
    W &= \om + \theta^\nu k_\nu + \frac12 \theta\bar\theta l,
  \end{aligned}
\end{equation}
where $\zeta^m(x)$, $\zeta^\mu(x)$ and $\om(x)$ are the parameters of
bosonic diffeomorphisms, supersymmetry transformations and Lorentz
boosts, respectively. In section \ref{sec:gauge-fixing} we made use of
the components to zero and first order in $\theta$ of (\ref{labl9}) to
argue that certain gauge conditions may be imposed. This, however,
does not mean that those components are fixed (e.\,g.\ to zero), but
only that the higher order components depend on the zero order ones
and on the fields constituting the supermultiplets. In order to
maintain the gauge conditions (\ref{egacof}) and the first relation
(\ref{egacot}) we must have
\begin{equation}
  \label{labl7}
  \dl E^{(0)}\_\al\^m = 0, \quad \dl E^{(0)}\_\al\^\mu = 0, \quad \dl
  \Conn^{(0)}\_\al = 0.
\end{equation}
In terms of the transformation components in (\ref{labl9}) one easily
finds from (\ref{labxx})
\begin{equation}
  \begin{aligned}
    k_\nu\^m &= i (\g^m \bar\zeta)_\nu, \\
    k_\nu\^\mu &= i (\g^b \bar\zeta)_\nu \chi_b\^\mu + \frac12 \om
    \g^5\_\nu\^\mu, \\
    k_\nu &= -A (\g^5 \bar\zeta)_\nu - i \rho_b (\g^b \bar\zeta)_\nu.
  \end{aligned}
\end{equation}

To be consistent with the remaining gauge conditions one has to solve
the equations
\begin{equation}
  \begin{aligned}
    \dl \left( E^{(1)}\_{\nu\al}\^m - E^{(1)}\_{\al\nu}\^m \right) &=
    0, \\
    \dl \left(E^{(1)}\_{\nu\al}\^\mu - E^{(1)}\_{\al\nu}\^\mu \right)
    &= 0, \\
    \dl \left( \Conn^{(1)}\_{\nu\al} - \Conn^{(1)}\_{\al\nu} \right)
    &= 0
  \end{aligned}
\end{equation}
which determine the remaining functions $l^m$, $l^\mu$, and $l$.  To
clarify the geometrical meaning of the transformations we write the
final answer separately for bosonic diffeomorphisms and Lorentz
rotations (setting $\z^\mu=0$)
\begin{align}                                        
  \xi^m &= \zeta^m, \label{ebodpa} \\
  \xi^\mu &= \frac12 \om  (\theta \g^5)^\mu, \label{efelpa} \\
  W &= \om, \label{ebolpa}
\end{align}
and for local supersymmetry transformations (setting $\z^m = 0$, $\om =
0$)
\begin{align}
  \xi^m &= i (\theta \g^m \bar\zeta) - \theta\bar\theta (\zeta
  \bar\lm^m), \label{ebospa} \\
  \xi^\mu &= \zeta^\mu + i (\theta \g^b \bar\zeta) \chi_b\^\mu +
  \frac12 \theta\bar\theta \left[ \lm^2 \zeta^\mu + \frac{i}{2} c_b
    (\zeta \g^b)^\mu \right], \label{eferpa} \\
  W &=- A (\theta \g^5 \bar\zeta) - i \rho_b (\theta \g^b \bar\zeta) -
  \theta\bar\theta \left[ \e^{mn} (\zeta \widetilde\nb_m \bar\chi_n) +
    c_b (\zeta \g^5 \bar\lm^b) - i \lm^2 (\zeta \g^5 \bar\chi)
  \right]. \label{elorpa}
\end{align}

We see that the bosonic vector field $\zeta^m(x)$ only enters as the
zeroth component of $\xi^m(z)$.  Notice that with respect to the
Lorentz boost the coordinate $\theta^\mu$ changes as if it were a
spinor in the anholonomic basis. This is a consequence of the `gauge'
condition (\ref{egacof}) which simply implies that in the zeroth order
of $\theta$ the spinor components in the holonomic and in the
anholonomic basis are identified and, therefore, must be transformed
simultaneously.  The local supersymmetry transformation parameters
$\zeta^\mu(x)$ appear in a complicated manner and produce nontrivial
transformations in the bosonic subspace as well as a local Lorentz
boost.  Notice also that the parameters only depend on the
supergravity multiplet $\cal E$. This means that consistent
supergravity can be and was constructed in terms of that quantity.
The transformation rules for the supergravity multiplet are obtained
by considering the variations $\dl E^{(0)}\_a\^m$, $\dl
E^{(0)}\_a\^\mu$, $\dl \Conn^{(1)}\_\al$ under the transformations
(\ref{ebodpa})--(\ref{elorpa}). For bosonic diffeomorphisms and
Lorentz boosts one verifies the desired transformation laws for fields
according to their representations
\begin{align}
  \dl e_a\^m &= \zeta^n \pl_n e_a\^m - e_a\^n \pl_n \zeta^m
  - \om \e_a\^b e_b\^m, \label{ezwetl} \\
  \dl \chi_a\^\mu &= \zeta^n \pl_n \chi_a\^\mu - \om \e_a\^b \chi_b\^\mu
  - \frac12 \om (\chi_a \g^5)^\mu,  \label{ershtl} \\
  \dl A &= \zeta^m \pl_m A. \label{elsbtl}
\end{align}
This a posteriori justifies the gauge fixing procedure in section
\ref{sec:gauge-fixing}, restricting general superdiffeomorphisms.
Under local supersymmetry the supergravity supermultiplet $\cal E$
transforms as
\begin{align}
  \dl e_a\^m &= 2i (\zeta \g^m \bar\chi_a), \label{ezwetr} \\
  \dl \chi_a\^\mu &= -\widetilde\nb_a \zeta^\mu - 2i (\chi \bar\lm_a)
  \zeta^\mu - 2i (\lm_b \bar\chi_a) (\zeta \g^b)^\mu - \frac{i}{2} A
  (\zeta \g_a)^\mu \label{ershtr} \\
  \dl A &= 2 \e^{mn} (\zeta \g^5 \widetilde\nb_m \bar\chi_n) + 2i
  \lm^2 (\z \bar\chi) - 2i A (\z \bar\chi). \label{erhftr}
\end{align}
Transformation rules for the connection supermultiplet $\Omega_a =
\{\om_a, u_a\^\mu, \upsilon_a\}$ may be read off from the variation of
$\Conn_a$. These rules appear to be rather complicated.  One of the
reasons for this is that this supermultiplet consists of connections
with more complicated transformational properties than tensors.  It
turns out that the introduction of a new torsion supermultiplet $ T_a
= \{\Tor_a, \tau_a\^\mu, s_a\}$ is more convenient than to work in
terms of $\Omega_a$. The field $\Tor_a$ is given by (\ref{estocz}),
whereas $\tau_a\^\mu$ and $s_a$ are the first and second order
components of the torsion superfield $T_a$ as defined by
(\ref{estocf}) and (\ref{estocs}). The new fields are related to the
old ones by algebraic invertible equations and thus both
supermultiplets are equivalent. It should be admitted that in terms of
$T_a$ the interaction with matter superfields is very likely to be
more complicated in higher orders of $\theta$ (because in the zeroth
order we still keep the Lorentz connection as an independent variable)
but this is beyond the scope of the present paper. In any case the
transformation rules of $\Omega$ can be deduced by the interested
reader from the formulas of the present paper.

To obtain the transformation rules for $T_a$ we consider the variation
\begin{displaymath}
  \dl T_a = \xi^m \pl_m T_a + \xi^\mu \pl_\mu T_a - W \e_a\^bT_b
\end{displaymath}
and arrive at the familiar transformations of bosonic diffeomorphisms
and Lorentz boosts
\begin{align}
  \dl\Tor_a &= \z^n \pl_n \Tor_a - \om \e_a\^b \Tor_b,
  \label{etroml} \\ 
  \dl\tau_a\^\mu &= \z^n \pl_n \tau_a\^\mu - \om \e_a\^b \tau_b\^\mu
  - \frac12 \om (\tau_a \g^5)^\mu, \label{etrftl} \\
  \dl s_a &= \z^n \pl_n s_a - \om \e_a\^b s_b. \label{etrbtl}
\end{align}
Under local supersymmetry $T_a$ transforms as
\begin{align}
  \dl\Tor_a &= (\z \bar\tau_a) \label{etrts0} \\
  \dl\tau_a\^\mu &= -i (\z \g^b)^\mu \left[ \widetilde\nb_b \Tor_a -
    4i \e_a\^c \Tor_c (\lm_b \g^5 \bar\chi) + (\chi_b \bar\tau_a)
  \right] \nonumber \\
  &\phantom{= M} - (\z \g^5)^\mu A \e_a\^b \Tor_b + \z^\mu s_a,
  \label{etrts1} \\
  \dl s_a &= -2 (\z \bar\lm^b) \widetilde\nb_b \Tor_a + 2 \e^{mn}
  (\z\widetilde \nb_m \bar\chi_n) \e_a\^b \Tor_b - 2i (\z \g^5
  \bar\chi) \lm^2 \e_a\^b t_b + i (\z \g^b \widetilde\nb_b \bar\tau_a)
  \nonumber \\
  &\phantom{= M} + (\z \bar \tau_a) \lm^2 + 4 \e_a\^b (\z \g^c
  \bar\tau_b) (\lm_c \g^5 \bar\chi) + A \e_a\^b (\z \g^5 \bar\tau_b) -
  2i (\z \bar\chi) s_a. \label{etrts2}
\end{align}
Because of its importance we also add the transformation of the
bosonic Lorentz connection, although it is implied by (\ref{etrts0}),
\begin{equation}                                       
  \label{etrom}
  \begin{split}
    \dl\om_a &= \e_a\^b (\z \bar\tau_b) + 2i \e^{mn} (\z \g_a
    \widetilde\nb_m \bar\chi_n) + 2i c_b (\z \g^b \g^5 \bar\chi_a) \\
    &\phantom{= M} + 8 (\z \g^5 \bar\lm_a) \chi^2 + 2 (\z \g_a \g^5
    \bar\chi) \lm^2 - 2 A (\z \g^5 \bar\lm_a).
  \end{split}
\end{equation}

We see that the parameter $\zeta^m(x)$ exactly produces the general
coordinate transformations (diffeomorphisms) of the bosonic subspace.
The Lorentz boost with parameter $\om(x)$ not only rotates the
anholonomic indices of vectors and spinors but also the holonomic
spinor indices.  Thus we have identified the Lorentz boost in the
tangent space with the Lorentz subgroup entering the group of general
coordinate transformations in superspace.  It seems remarkable that
also for our generalized supergravity the transformation of the
supergravity supermultiplet does not involve extra fields and remains
the same as for vanishing bosonic torsion \cite{howe79}. The Lorentz
superconnection $\Conn_a$ as a superfield or the trace of the
supertorsion $T_a$ form separate supermultiplets whose transformation
rules contain the supergravity multiplet. The same is known to happen
if one adds additional matter superfields \cite{ikeda94}. Then the
transformation of its components will involve the supergravity
supermultiplet because it explicitly enters the transformation
parameters (\ref{ebospa})--(\ref{elorpa}).


\section{Supergravity Lagrangians}
\label{sec:lagrangians}

The construction of generic supergravity Lagrangians within the
superspace approach is simple but implies lengthy calculations when
one desires to write them in terms of all the fields contained in the
supermultiplets. A functional
\begin{equation}
  \label{esupac}
  I = \int d^2x d^2\theta\; E\; L(x, \theta),
\end{equation}
($E = \sdet E_M\^A$ denotes the Berezinian (superdeterminant) of the
supervierbein \cite{berezin66,dewitt84} and $L$ is an arbitrary scalar
superfield built from supermultiplets) after integration over $\theta$
yields a supergravity model written for a set of fields over
two-dimensional space-time. If a supermanifold has a boundary or a
nontrivial topology then the usual integration rule over $\theta$ must
be modified \cite{supergeometry}. In our case we have two
supermultiplets united in two superfields $S$ and $T_a$. The
anholonomic indices are transformed only under a Lorentz boost, and
Latin and Greek indices do not mix. Thus one may construct scalar
Lagrangians by contracting Latin and Greek indices separately. There
are, of course, an infinite number of choices. For example, using the
definition of the scalar curvature of superspace (\ref{esuscu}) and
the explicit form of its components (\ref{ecu1})--(\ref{ecu3}) one
easily finds for the supercurvature invariant (\ref{esuscu})
\begin{equation}
  \label{esscsu}
  R = \e^{\al\bt} \nb_\bt \nb_\al S - 2 S^2 + 2 \nb_a T^a - 2 T_a T^a
  - 2 S.
\end{equation}
It is important to note that any term on the right hand side is a
scalar superfield and can be chosen as a Lagrangian.

Let us consider some of the simplest Lagrangians, to be obtained by
special choices of $L$ in (\ref{esupac}). Using the definition

\begin{equation}
  \label{esudet}         
  E = \sdet E_M\^A = \frac{\det(E_m\^a - E_m\^\bt E^{-1}\_\bt\^\nu
    E_\nu\^a)}{\det E_\mu\^\al}
\end{equation}
and the expression for the supervierbein in terms of the supergravity
supermultiplet (\ref{esuvbb})--(\ref{esuvff}) we have
\begin{equation}
  \label{esudec}
  E = \det e_m\^a \left[ 1 + 2i (\theta \bar\chi) + \frac12
    \theta\bar\theta (2 \chi^2 + \lm^2 + A) \right].
\end{equation}

Integrating $\theta$ in (\ref{esupac}) produces the second order
component of the product of the superdeterminant and the scalar
superfield chosen as a Lagrangian.  The simplest example $L=1$ would
correspond to the cosmological constant in ordinary gravity. In
supergravity, with the superdeterminant alone, we obtain
\begin{equation}
  \label{ecocol}
  {\cal L}_1 = A + 2 \chi^2 + \lm^2.
\end{equation}
Here and below in such Lagrangians in 2d space-time we drop the factor
$e = \det e_m\^a$ for brevity, thus $I_1 = \int d^2\!x \, e \, {\cal
  L}_1$.  This Lagrangian by itself has only trivial solutions but may
yield nontrivial contributions if added to other Lagrangians.  We see
that no cosmological constant can be added to a supergravity model in
this way.

For $ L = S$ one arrives at
\begin{equation}
  \label{elagrs}
  {\cal L}_S = -\frac12 \widetilde R - 4i \widetilde\nb_a (\lm^a
  \bar\chi).
\end{equation}
where $\widetilde R$ and $\widetilde \nabla$ are defined in appendix
\ref{app:anhol-basis}.  Thus this Lagrangian multiplied by $e$ equals
to a total derivative.  Therefore, the `minimal' supergravity in two
dimensions is as trivial as the bosonic Hilbert-Einstein action,
represented by the first term in (\ref{elagrs}). However, the
components of the supergravity multiplet are essential for
constructing interactions with matter superfields.  For example, they
are of particular relevance to superstring theory.

Other scalar superfields provide nontrivial models with second order
equations of motion.  For $ L = S^2$ we have
\begin{equation}
  \label{elagss}
  \begin{split}
    {\cal L}_{S^2} &= -A \widetilde R - A^3 + 4 \e^{mn} \e^{pr}
    ((\widetilde\nb_m \chi_n) (\widetilde\nb_p \bar\chi_r)) + 8i (A -
    \lm^2) \e^{mn} (\chi \g^5 \widetilde\nb_m \bar\chi_n) \\
    &\phantom{= M} - 8i A \widetilde\nb_a (\lm^a \bar\chi) - 8 A
    \chi^2 \lm^2 - A^2 (\chi^a \bar\chi_a).
  \end{split}
\end{equation}
The term $\nb^\al\nb_\al S$ is also contained in the expression
(\ref{esscsu}) for $R$. Taken as a Lagrangian $L$ the related ${\cal
  L}_{\nabla^2 S}$ turns out to be proportional to (\ref{elagss}) up
to a total divergence.

Eq.\ (\ref{elagss}) represents a supergravity theory with vanishing
bosonic torsion. Therefore, we could have obtained it without the
extension discussed in our present paper. The Rarita-Schwinger field
appears with first and second derivatives. On the other hand, the
scalar field $A$ is nondynamical. Due to the linear term $A\widetilde
R$, however, by the conformal transformation $e_m\^a \to A e_m\^a$ of
the zweibein a kinetic term for $A$ may be produced, and $-(\ln \vert
A\vert) / 2 = \phi$ may be interpreted as a dilaton field
\cite{general-dilaton-gravity}. In that case from the bosonic part of
(\ref{elagss}) alone theories with interesting highly nontrivial
singularity properties may be obtained. A different approach to
supersymmetric dilaton gravity is adopted in
\cite{dilaton-supergravity,ikeda94}, where an extra dilaton superfield
is introduced. In our case the bosonic dilaton field arises from the
scalar component of the gravity supermultiplet.

In a similar way one obtains the two simplest Lagrangians containing
the torsion supermultiplet $T_a$ with at most second order
e.\,o.\,m.\,-s.  $L = T^2$ in (\ref{esupac}) leads to (cf.\ 
(\ref{estocz}))
\begin{equation}
  \label{elagtt}
  \begin{split}
    {\cal L}_{T^2} &= (A + 2 \chi^2 + \lm^2) \tor_a \tor^a + 2 s^a
    \Tor_a - (\tau^a \bar\tau_a) \\
    &\phantom{= M} - 4i (\chi \bar\tau^a)\, \Tor_a - 8i A \tor^a
    (\lm_a \bar\chi) + 8 A \chi^2 \lm^2.
  \end{split}
\end{equation}                                      
The Lagrangian for $\nb_a T^a$ turns out to be the same up to a total
divergence.  Therefore, their difference does not contribute to the
analog of the Hilbert-Einstein action $L_R$ in superspace, constructed
with (\ref{esscsu}) as a Lagrangian. Thus this entire Lagrangian in
superspace is proportional to (\ref{elagss}) only. The Lagrangian
(\ref{elagtt}) leads to the constraint $T_a = 0$, i.\,e.\ the
Lagrangians (\ref{ecocol})--(\ref{elagss}) exhaust the set of
nontrivial ones with not more than two derivatives in the
e.\,o.\,m.\,-s for the fields $e_a\^m$, $\om_a$ and their supergravity
partners. We are thus led to the conclusion that in any construction
of $L$ using the superfields $S$ and $T_a$ and requiring at most first
derivatives of the supergravity Cartan variables in ${\cal L}$,
bosonic torsion has to vanish after all.

The construction of the supergravity models above is based on the
action principle in superspace. Another possibility to get a
supersymmetric extension of a given bosonic model consists in the
generalization of the equations of motion to superspace. For example,
the super Liouville model can be naturally formulated in terms of the
supergravity multiplet. It is well known that two-dimensional constant
curvature gravity \cite{const-curvature-gravity}, $\widetilde R =
\text{const}$, in the conformal gauge reduces to the Liouville
equation.  The super extension of this model is given by the invariant
equation
\begin{displaymath}
  S = C = \text{const},
\end{displaymath}
or in components
\begin{align*}
  A &= C, \\
  \e^{mn} (\g^5 \widetilde\nabla_m \bar\chi_n)_\al + i \lm^2
  \bar\chi_\al - i C \bar\chi_\al &= 0, \\
  \frac12 \widetilde R - 4i \e^{mn} (\chi \g^5 \widetilde\nabla_m
  \bar\chi_n) + 4i \widetilde\nabla_a (\lm^a \bar\chi) + 4\chi^2 \lm^2
  + C (\chi^a \bar\chi_a) + C^2 &= 0.
\end{align*}
The field $A$ is constant due to the first equation, and
the last two of this supercovariant system of equations may be
rewritten in equivalent form
\begin{align}
  -(\g^a \widetilde\nabla_a \bar\chi)_\al - \widetilde\nabla_a
  \bar\lm^a\_\al + i \lm^2 \bar\chi_\al - i C \bar\chi_\al &= 0,
  \label{esulio} \\ 
  \widetilde R + 8i \widetilde\nabla_a (\lm^a \bar\chi) + 2 C (2
  \chi^2 + \lm^2) + 2 C^2 &= 0. \label{esulit}
\end{align}
These equations reduce to the constant curvature gravity in the
absence of the Rarita-Schwinger field. It seems to be the simplest
nontrivial supergravity model in two dimensions.  This super extension
of the Liouville model differs from the known generalizations. The
effective superspace action for a superstring off the critical
dimension may be also considered as the super extension of the
Liouville model \cite{super-liouville1}. In that case the action
depends on the extra scalar superfield. Another super extension of the
Liouville model may be found in \cite{super-liouville2} where the
Liouville action is generalized without incorporating the general
coordinate invariance.

In order to explore possibilities of some higher order Lagrangians it
is sufficient for a first orientation to determine their bosonic
parts. Of course, the supplementing superfields can be introduced in
all cases in a straightforward manner.

For vanishing fermionic fields the inverse supervielbein
(\ref{einsvf})--(\ref{einsvg}) and the Lorentz superconnection
(\ref{esulob}), (\ref{esulof}) take a particularly simple form
\begin{align}
  E_a\^m &= e_a\^m, \label{einbvf} \\
  E_a\^\mu &= -\frac{i}{2} A (\theta \g_a)^\mu + \frac12
  \widetilde\om_a (\theta \g^5)^\mu \label{einbvs} \\
  E_\al\^m &= i (\g^m \bar\theta)_\al, \label{einbvt} \\
  E_\al\^\mu &= \dl_\al\^\mu - \frac12 \theta\bar\theta \left[ \frac12
    A \dl_\al\^\mu + \frac{i}{2} c_b \g^b\_\al\^\mu \right],
  \label{einbvu} \\
  \Conn_a &= \om_a + \frac12 \theta\bar\theta \e_a\^b (s_b + \pl_b A),
  \label{esblob} \\
  \Conn_\al &= A (\theta \g^5 \e)_\al - i \widetilde\om_b (\theta \g^b
  \e)_\al. \label{esblof}
\end{align}
and the corresponding supervielbein (\ref{esuvbb})--(\ref{esuvff})
becomes
\begin{align}
  E_m\^a &= e_m\^a + \frac12 \theta\bar\theta A e_m\^a, \label{esbvbb}
  \\
  E_m\^\al &= \frac{i}{2} A (\theta \g_m)^\al -
  \frac12 \widetilde\om_m (\theta \g^5)^\al, \label{esbvbf} \\
  E_\mu\^a &= i (\theta \g^a \e)_\mu, \label{esbvfb} \\
  E_\mu\^\al &= \dl_\mu\^\al \left( 1 - \frac14 \theta\bar\theta A
  \right). \label{esbvff}
\end{align}
Also the superfields $S$ and $T_a$ simplify greatly:
\begin{align}
  S &= A- \frac12 \theta\bar\theta \left[ \frac12 \widetilde R + A^2
  \right], \label{escsbf} \\
  T_a &= t_a + \frac12 \theta\bar\theta s_a. \label{esutbr}
\end{align}

Together with the superdeterminant
\begin{equation}
  \label{esudeb}
  E = e \left( 1 + \frac12 \theta\bar\theta A \right)
\end{equation}
the bosonic parts of the possible supergravity Lagrangians
(\ref{esupac}), constructed from scalar and vector superfields
(\ref{escsbf}) and (\ref{esutbr}) are determined easily and can be
generalized to obtain other models which permit a direct
'dilatonization'.  Consider an arbitrary power $k$ of the scalar
superfield $S^k$. The corresponding bosonic Lagrangian is
\begin{displaymath}
  {\cal L}_{S^k} = - \frac{k}{2} A^{k-1} \widetilde R - (k - 1)
  A^{k+1}.
\end{displaymath}
For negative scalar curvature the field $A$ can be eliminated using
its equation of motion, and the Lagrangian becomes
\begin{displaymath}
  {\cal L}_{S^k} \approx (-\widetilde R)^{\frac{k+1}{2}}.
\end{displaymath}
This Lagrangian yields a large nontrivial class of gravity models in
two dimensions. All of them are integrable
\cite{general-dilaton-gravity} (that is one can write down a general
solution to the equations of motion and even analyse the corresponding
unique topology of the space-time) and can be made locally
supersymmetric. Identifying now $kA^{k-1} = -\exp(-2\phi)/2$, again
$A$ (or $\phi$) can be made dynamical by a conformal transformation of
$e_m\^a$ as in the case $k = 1$.  Of course, also polynomials and even
arbitrary functions of $S$ could be considered in $L$
\cite{hindawi-ovrut-waldram96}. Also conformal transformations by a
superfield may provide a supersymmetric 'dilatonization'
\cite{hindawi-ovrut-waldram96}.  The bosonic part of those theories
then becomes just a generic one of covariant PSM-models
\cite{non-einsteinian-gravity} with vanishing torsion.  Supplementing
the supersymmetric part immediately provides again supergravity
extensions.

Let us briefly discuss other possible generalizations. In the examples
above the scalar field $A$ acted essentially as an auxiliary field
because it entered the Lagrangian without kinetic term and could be
eliminated by solving its algebraic equation of motion. However, in
general, also a supersymmetric kinetic term for $A$ may exist.
Consider, for example, the scalar superfield $\nb^\al S\nb_\al S$. The
corresponding Lagrangian has the form
\begin{displaymath}
  {\cal L}_{\nb^\al S \nb_\al S} = 2 \et^{ab} \pl_a A \pl_b A +
  \frac12 (\widetilde R + 2 A^2)^2.
\end{displaymath}
The appearance of the (torsionless) scalar curvature in the second
power yields higher derivatives.  The other possible candidate with
second derivatives for $A$ is
\begin{displaymath}
  {\cal L}_{\nb^a S \nb_a S} = -3 A \pl^a A \pl_a A - \pl^a A \pl_a
  \widetilde R.
\end{displaymath}

Another example, involving nonvanishing bosonic torsion, is the (for
the zweibein only) higher derivative Lagrangian containing a kinetic
term for the Lorentz connection $\om_a$
\begin{displaymath}
  {\cal L}_{\nb^\al T^a \nb_\al T_a} = 2 \widetilde\nb^a \tor^b
  \widetilde\nb_a \tor_b + 2 s^a s_a + 2 A^2 \tor^a \tor_a.
\end{displaymath}
Such a kinetic term for the Lorentz connection $\om_a$ is also
contained in a Lagrangian from $L = R^2$. For simplicity we
abbreviate the scalar supercurvature by extracting the last term in
(\ref{esscsu})
\begin{displaymath}
  \underline R = -\e^{ab} F_{ba} = R + 2 S.
\end{displaymath}
A simple calculation gives the bosonic Lagrangian (quantities with
hats are built with the $x$-space connection $\om_m$, cf.\ appendix
\ref{app:anhol-basis})
\begin{displaymath}
  {\cal L}_{\underline R\^2} = 3 A \widehat R^2 + 4 \widehat R
  \widehat\nb^a \widehat\nb_a A + 4 A^3 \widehat R + 4 \widetilde\nb_a
  s^a + 8 \tor^a \pl_a A.
\end{displaymath}

There is, of course, an infinite number of other higher order bosonic
Lagrangians which can be made locally supersymmetric, also including
higher powers of torsion.


\section{Summary and Outlook}
\label{sec:summary}

The generalization of rigid supersymmetry to generalized theories of
supergravity does not necessitate the validity of the bosonic torsion
constraint. Dropping the latter in 1+1 space-time we solve the minimal
set of constraints (\ref{etorco}) for $N = (1,1)$ superspace. The
computational problems which would occur following the approach of the
seminal work by Howe \cite{howe79} are greatly reduced by working in
terms of the \emph{inverse} supervierbein and the Lorentz
superconnection.  After conventional gauge fixing of
superdiffeomorphisms and Lorentz boosts it is possible to \emph{first}
solve the constraints.  The inverse supervierbein turns out to be
expressed only in terms of a supergravity multiplet ${\cal E} =
(e_a\^m, \chi_a\^\mu, A)$ consisting of the zweibein, a
Rarita-Schwinger field and a spinor field contained in the (reducible)
field $\chi_a\^\mu$ and a scalar $A$. The Lorentz superconnection in
turn, beside the components of $\cal E$ has arbitrary components
expressible either as a supermultiplet of connections $\Omega_a =
(\om_a, u_a\^\mu, \upsilon_a)$ or, alternatively, by a torsion
multiplet $T_a$ of the same structure. As a consequence of our
approach, the Bianchi identities are fulfilled identically.
Nevertheless, they are very useful in order to find out how the
components of ${\cal E}$ may be summarized in a scalar superfield $S$;
$T_a$ may be interpreted as the trace of the supertorsion. Since the
component fields are found to transform with respect to the correct
local diffeomorphisms, local Lorentz and supersymmetry
transformations, a generic superfield Lagrangian is a superscalar
built from $S$ and $T_a$, multiplied by the superdeterminant of
$E_M\^A$. Considering the simplest example which leads to at most
second order equations of motion for the fields we find that
nontrivial Lagrangians then are restricted to the case $T_a = 0$.
Therefore no action of this type may produce the immediate
generalization of two-dimensional gravity with torsion
\cite{dynamic-torsion}.  However, 2d theories with arbitrary powers in
bosonic curvature \cite{hindawi-ovrut-waldram96} --- and vanishing
torsion --- in our approach are readily extended to completely
supersymmetric ones with all superfield-partners included.  Such
theories also allow 'dilatonization', i.\,e.\ by conformal
transformation of the zweibein or metric involving the ---
nondynamical --- scalar field $A$ in $S$, globally quite different
dilaton theories with $A$ related to the dilaton field may be produced
\cite{general-dilaton-gravity}.  There are even many types of higher
derivative theories with nonvanishing torsion ($T_a \neq 0$) to be
exploited in further work. From recent results on general classical
and quantum solutions for any such theory in the purely bosonic case
\cite{non-einsteinian-gravity,general-dilaton-gravity,kummer-liebl-vassilevich97}
it may be conjectured that their integrability may be extended to
these supergravity generalizations, allowing for exactly solvable
models for black holes including matter in the form of superpartners
of the zweibein and the Lorentz connection. Also the relation to the
first order formulation in
\cite{non-einsteinian-gravity,general-dilaton-gravity,kummer-liebl-vassilevich97}
needs clarification. An interesting generalisation to the heterotic
supergeometry \cite{heterotic-supergeometry} may be relevant to
heterotic string theory. These are some of the topics we intend to
tackle in further work.

\begin{appendix}


\section{Conventions of Spinor-Space and Spinors}
\label{app:spinors}
  
Of course, the properties of Clifford algebras and spinors in any
number of dimensions (including $d = 1+1$) are well-known, but in view
of the tedious calculations required in our present work we include
this appendix in order to prevent any misunderstandings of our results
and facilitate the task of the intrepid reader who wants to redo
derivations.
  
The $\g$ matrices which are the elements of the Clifford algebra
defined by the relation
\begin{equation}
  \label{edefgm}
  \g^a \g^b + \g^b \g^a =2 \et^{ab}, \quad \et_{ab} = \et^{ab} =
  \diag(+-),
\end{equation}
are represented by two-dimensional matrices
\begin{equation}
  \label{egamma}
  \g^0\_\al\^\bt = \mtrx{0}{1}{1}{0}, \quad \g^1\_\al\^\bt =
  \mtrx{0}{1}{-1}{0}.
\end{equation}
As indicated, the lower index is assumed to be the first one.  The
spinor indices are often suppressed assuming the summation from `ten
to four'.  The generator of a Lorentz boost (hyperbolic rotation) has
the form
\begin{equation}
  \sigma^{ab} = \frac12 (\g^a \g^b - \g^b \g^a) = \e^{ab} \g^5,
\end{equation}
where
\begin{equation}
  \g^5 = -\g^0 \g^1 = \mtrx{1}{0}{0}{-1}, \quad (\g^5)^2 = 1,
\end{equation}
and
\begin{equation}
  \e_{ab} = -\e^{ab} = \mtrx{0}{1}{-1}{0}
\end{equation}
is the totally antisymmetric tensor with vector indices obeying
\begin{equation}
  \e^{ab} \e_{cd} = -\dl_c\^a \dl_d\^b + \dl_c\^b \dl_d\^a , \quad
  \e^{ab} \e_{bd} = \dl_d\^a, \quad \e^{ab} \e_{ba} = 2,
\end{equation}
where $\dl_a\^b = \dl_a^b$ denotes the Kronecker symbol. In two
dimensions the $\g$-matrices satisfy the relation
\begin{equation}
  \label{egmpro1}
  \g^a \g^b = \et^{ab} + \e^{ab} \g^5,
\end{equation}
which is equivalent to the definition (\ref{edefgm}). The following
formulas are frequently used in our calculations:
\begin{equation}
  \label{egmpro2}
  \begin{aligned}[t]
    \g^a \g_a &=2, \\
    \g^a \g^5 + \g^5 \g^a &= 0, \\
    \tr(\g^a \g^b) &= 2 \et^{ab}.
  \end{aligned} \quad
  \begin{aligned}[t]
    \g^a \g^b \g_a &= 0, \\
    \g^a \g^5 &= \g^b \e_b\^a,
  \end{aligned}
\end{equation}
As usual the trace of the product of an odd number of $\g$-matrices
vanishes.

In two dimensions the $\g$-matrices satisfy the Fierz identity
\begin{multline}
  \label{efierz}
  2 \g^a\_\al\^\g \g^b\_\bt\^\dl = \g^a\_\al\^\dl
  \g^b\_\bt\^\g + \g^b\_\al\^\dl \g^a\_\bt\^\g + \\
  + \eta^{ab} (\dl_\al\^\dl \dl_\bt\^\g - \g^5\_\al\^\dl \g^5\_\bt\^\g
  - \g^c\_\al\^\dl \g_{c\bt}\^\g) + \e^{ab} (\g^5\_\al\^\dl
  \dl_\bt\^\g - \dl_\al\^\dl \g^5\_\bt\^\g),
\end{multline}
which can be checked by direct calculation. Different contractions of
it with $\g$-matrices then yield different but equivalent versions
\begin{align}
  2 \dl_\al\^\g \dl_\bt\^\dl &= \dl_\al\^\dl \dl_\bt\^\g
  +\g^5\_\al\^\dl \g^5\_\bt\^\g + \g^a\_\al\^\dl \g_{a\bt}\^\g,
  \label{efier1} \\
  2 \g^5\_\al\^\g \g^5\_\bt\^\dl &= \dl_\al\^\dl \dl_\bt\^\g +
  \g^5\_\al\^\dl \g^5\_\bt\^\g - \g^a\_\al\^\dl \g_{a\bt}\^\g,
  \label{efier2} \\
  \g^a\_\al\^\g \g_{a\bt}\^\dl &= \dl_\al\^\dl \dl_\bt\^\g -
  \g^5\_\al\^\dl \g^5\_\bt\^\g, \label{efier3}
\end{align}
which allow to manipulate third and higher order monomials in spinors.
Notice that equation (\ref{egmpro1}) is also the consequence of
(\ref{efierz}).

The totally antisymmetric tensor and the Minkowskian metric satisfy
the Fierz-type identity in two dimensions
\begin{equation}
  \eta_{ab} \e_{cd} + \eta_{da} \e_{bc} + \eta_{cd} \e_{ab} +
  \eta_{bc} \e_{da} = 0,
\end{equation}
which allows to make rearrangements in third and higher order
monomials of Lorentz vectors.

A Dirac spinor in two dimensions, forming an irreducible
representation for the full Lorentz group including space and time
reflections, has two complex components. We write it --- in contrast
to the usual convention in field theory, but in agreement with
conventional superspace notations --- as a row
\begin{equation}
  \psi^\al = (\psi^1, \psi^2).
\end{equation}
In our notation the first and second components of a Dirac spinor
correspond to right and left chiral Weyl spinors $\psi^{(\pm)}$,
respectively,
\begin{alignat}{2}
  \psi^{(\pm)} &= \psi \frac{1 \pm \g^5}{2}, &\quad
  \psi^{(\pm)} \g^5 &= \pm \psi^{(\pm)}, \\
  \intertext{where} \psi^{(+)} &= (\psi^1, 0), &\quad \psi^{(-)} &=
  (0, \psi^2).
\end{alignat}
Here matrices act on spinors form the right according to the usual
multiplication law. All spinors are always assumed to be anticommuting
variables.  The notation with upper indices is a consequence of our
convention to contract indices, together with the usual multiplication
rule for matrices. Under the Lorentz boost by the parameter $\om$
spinors transform as
\begin{equation}
  \psi'\^\al = \psi^\bt S_\bt\^\al,
\end{equation}
where
\begin{equation}
  \label{erotsp}
  S_\bt\^\al = \dl_\bt\^\al \cosh{\frac{\om}{2}} - \g^5\_\bt\^\al
  \sinh{\frac{\om}{2}} = \mtrx{e^{-\om/2}}{0}{0}{e^{+\om/2}},
\end{equation}
when the Lorentz boost of a vector is given by the matrix
\begin{equation}
  \label{erotve}
  S_b\^a = \dl_b\^a \cosh{\om} + \e_b\^a \sinh{\om} =
  \mtrx{\cosh{\om}}{-\sinh{\om}}{-\sinh{\om}}{\cosh{\om}}.
\end{equation}
By (\ref{erotsp}), (\ref{erotve}) the $\g$-matrices are invariant
under simultaneous transformation of Latin and Greek indices. This
requirement fixes the relative factors in the bosonic and fermionic
sectors of the Lorentz generator (\ref{eantis}).

Dirac conjugation is defined in the usual way and is written as a
column
\begin{equation}
  \label{edirco}
  \bar\psi_\al = \psi^{\bt*} g_{\bt\al} = \clmn{\psi^{2*}}{\psi^{1*}},
  \quad g_{\bt\al} = \mtrx{0}{1}{1}{0},
\end{equation}
where star $*$ denotes complex conjugation. Here we see that the role
of the matrix $\g^0$ is twofold. First, it is an operator in the
spinor space and thus has one lower and one upper index
(\ref{egamma}). The same matrix defines the metric in spinor space.
Therefore it is written in (\ref{edirco}) with two lower indices and
for clarity is denoted by the different symbol $g_{\al\bt}$.

In our conventions bilinear forms of spinors $\psi$ and $\h$ appear as
\begin{equation}
  \psi \G \bar\h = \psi^\al \G_\al\^\bt \bar\h_\bt,
\end{equation}
where $\G$ denotes any polynomial of the unit and the $\g$-matrices.
Under a Lorentz boost the bilinear forms $\psi \g^a \bar\h$ and $\psi
\bar\h$, $\psi \g^5 \bar\h$ transform as a vector and as scalars.

Among the discrete transformations the parity transformation (supposed
to be linear)
\begin{equation}
  P \maps \psi^\al \to \psi_p^\al = \psi^\bt P_\bt\^\al
\end{equation}
is uniquely defined by
\begin{equation}
  \begin{aligned}
    \psi_p \bar\h_p &= \psi \bar\h, \\
    \psi_p \g^0 \bar\h_p &= \psi \g^0 \bar\h, \\
    \psi_p \g^1 \bar\h_p &= -\psi \g^1 \bar\h,
  \end{aligned}
\end{equation}
up to an arbitrary complex number with unit modulus. We choose it to
be real
\begin{equation}
  P = \g^0.
\end{equation}
It can be checked easily that
\begin{equation}
  \psi_p \g^5 \bar\h_p = -\psi \g^5 \bar\h,
\end{equation}
i.\,e.\ is a pseudoscalar.


By definition charge (or Majorana) conjugation relates a spinor to its
Dirac conjugate
\begin{align}
  C \maps \psi^\al \to \psi_c^\al &= C^{\al\bt} \bar\psi_\bt, \\
  \bar\psi_{c\al} &= C_{\al\bt}^{-1} \psi^\bt.
\end{align}
The last equation is needed in order that the square of charge
conjugation results in the identical transformation. Because Dirac
conjugation is already defined by (\ref{edirco}) the charge
conjugation matrix must satisfy the relation
\begin{equation}
  \label{echarp}
  C_{\al\bt}^{-1} = C^{\dagger\g\dl} g_{\dl\al} g_{\g\bt},
\end{equation}
where the cross denotes hermitian conjugation.  Requiring
\begin{align}
  \psi_c \bar\psi_c &= \psi \bar\psi, \label{echars} \\
  \psi_c \g^a \bar\psi_c &= -\psi \g^a \bar\psi, \label{echarv}
\end{align}
in order to preserve the sign of mass and to invert the sign of the
electric charge, already equation (\ref{echars}) defines the charge
conjugation matrix up to an arbitrary complex number of unit modulus
which we fix to unity
\begin{equation}
  \label{echarep}
  C^{\al\bt} = \e^{\al\bt}, \quad C_{\al\bt}^{-1} = -\e_{\al\bt},
\end{equation}
where
\begin{equation}
  \label{eantit}
  \e_{\al\bt} = \e^{\al\bt} = \mtrx{0}{1}{-1}{0}
\end{equation}
is the totally antisymmetric tensor with spinor indices. It has the
properties
\begin{equation}
  \e^{\al\bt} \e_{\g\dl} = \dl_\g\^\al \dl_\dl\^\bt - \dl_\g\^\bt
  \dl_\dl\^\al , \quad \e^{\al\bt} \e_{\bt\g} = -\dl_\g\^\al, \quad
  \e^{\al\bt} \e_{\bt\al} = -2.
\end{equation}
With (\ref{echarep}) relations (\ref{echarp}), (\ref{echarv}) are
verified. Under charge conjugation a pseudoscalar changes its sign
\begin{equation}
  \psi_c \g^5 \bar\psi_c = -\psi \g^5 \bar\psi.
\end{equation}
In the Majorana spinor --- defined by the requirement that its Dirac
conjugate equals the charge conjugate ---
\begin{equation}
  \label{emajsp}
  \bar\psi_\al = C_{\al\bt}^{-1} \psi^\bt \quad \Leftrightarrow \quad
  \psi^{*\al} = \psi^\bt \g^5\_\bt\^\al
\end{equation}
the first component is real while the second is purely imaginary
\begin{equation}
  \psi^{1*} = \psi^1, \quad \psi^{2*} = -\psi^2.
\end{equation}

For Majorana spinors the bilinear form
\begin{equation}
  \psi \bar\chi = \psi^\al \chi^\bt \e_{\bt\al},
\end{equation}
can be considered to be defined by the metric $\e_{\al\bt}$ in spinor
space. It has no definite parity because the definition of a Majorana
spinor (\ref{emajsp}) implies the $\g^5$ matrix. The bilinear
combinations of Majorana spinors have the properties
\begin{displaymath}
  \begin{array}{rcccrcl}
    \psi\bar\chi&=&\chi\bar\psi&=&-\psi^1\chi^2+\psi^2\chi^1 & \quad
    &{\rm real}
    \\
    \psi\g^5\bar\chi&=&-\chi\g^5\bar\psi&=&-\psi^1\chi^2-\psi^2\chi^1
    & & {\rm real}
    \\
    \psi\g^0\bar\chi&=&-\chi\g^0\bar\psi&=&\psi^1\chi^1-\psi^2\chi^2 &
    &{\rm imaginary}
    \\
    \psi\g^1\bar\chi&=&-\chi\g^1\bar\psi&=&\psi^1\chi^1+\psi^2\chi^2 &
    & {\rm imaginary},
  \end{array}  
\end{displaymath}
where one should remember that complex conjugation changes the order
of anticommuting variables. In particular, for Majorana spinors
\begin{equation}
  \psi \g^a \bar\psi = 0, \quad \psi \g^5 \bar\psi = 0
\end{equation}
the only nonvanishing quadratic form being $\psi \bar\psi = \psi^\al
\psi^\beta \e_{\bt\al}$. As a consequence of equation (\ref{egmpro1})
a quadratic form in Majorana spinors with an arbitrary odd number of
$\g$-matrices is always zero.

The field $\chi^{a\al}$ has one vector and one spinor index. We assume
that for each $a$ it is a Majorana spinor.  Therefore it has two real
and two purely imaginary components forming a reducible representation
of the Lorentz group. In many applications it becomes extremely useful
to work with its Lorentz covariant decomposition
\begin{equation}
  \label{ederas}
  \chi^a = \chi\g^a + \lm^a,
\end{equation}
where
\begin{equation}
  \chi = \frac12 \chi^a \g_a, \quad \lm^a = \frac12 \chi^b \g^a \g_b.
\end{equation}
The spinor $\chi^\al$ and the spin-vector $\lm_a\^\al$ form
irreducible representations of the Lorentz group and each of them has
two independent components. The spin-vector $\lm_a$ satisfies the
Rarita-Schwinger condition
\begin{equation}
  \label{equlam}
  \lm_a \g^a = 0
\end{equation}
valid for such a field. In two dimensions equation (\ref{equlam}) may
be written in equivalent forms
\begin{equation}
  \label{eculef}
  \lm_a \g_b = \lm_b \g_a \quad \text{or} \quad \e^{ab} \lm_a \g_b =
  0.
\end{equation}
If one chooses the $\lm^{0\al}$ components as independent ones then
the components of $\lm^{1\al}$ can be found from (\ref{equlam}) to be
\begin{displaymath}
  \lm^{0\al} = (\lm^{01}, \lm^{02}), \quad \lm^{1\al} = (\lm^{01},
  -\lm^{02}).
\end{displaymath}
It is important to note that as a consequence any cubic or higher
monomial of the (anticommuting) $\chi$ or $\lm_a$ vanishes
identically. The field $\lm_a$ satisfies further the useful relation
\begin{equation}
  \e_a\^b \lm_b = \lm_a \g^5
\end{equation}
which together with (\ref{egmpro2}) yields
\begin{equation}
  \begin{aligned}
    \e_a\^b \chi_b &= -\chi \g_a \g^5 + \lm_a \g^5.
  \end{aligned}
\end{equation} 
For the sake of brevity we often introduce the obvious notations
\begin{equation}
  \chi^2 = \chi \bar\chi, \quad \lm^2 = \lm^a \bar\lm_a.
\end{equation}
Other convenient identities used for $\lambda_a$ in our present work
are
\begin{equation}
  \begin{aligned}
    (\lm_a \bar\lm_b) &= \frac12 \et_{ab} \lm^2, \\
    (\lm_a \g^5 \bar\lm_b) &= \frac12 \e_{ab} \lm^2, \\
    (\lm_a \g_c \bar\lm_b) &= 0.
  \end{aligned}
\end{equation}
The first of these identities can be proved by inserting the unit
matrix $\g_a\g^a/2$ inside the product and interchanging the indices
due to equation (\ref{eculef}).  The second and third equation is
antisymmetric in indices $a$, $b$, and therefore to be calculated
easily because they are proportional to $\e_{ab}$.

Quadratic combinations of the vector-spinor field can be decomposed in
terms of irreducible components:
\begin{equation}
  \begin{aligned}
    (\chi_a \bar\chi_b) &= \et_{ab} \left( -\chi^2 + \frac12 \lm^2
    \right) + 2 (\chi \g_a \bar\lm_b) \\
    (\chi_a \g^5 \bar\chi_b) &= \e_{ab} \left( \chi^2 + \frac12 \lm^2
    \right) \\
    (\chi_a \g_c \bar\chi_b) &= 2 \e_{ab} (\chi \g^5 \bar\lm_c) \\
    (\chi_a \g_c \g^5 \bar\chi_b) &= 2 \e_{ab} (\chi \bar\lm_c) \\
  \end{aligned}  
\end{equation}


\section{Anholonomic Basis in Two Dimensions}
\label{app:anhol-basis}

Purely bosonic two-dimensional space-time is best described in terms
of the anholonomic orthonormal basis
\begin{equation}
  \label{eanhol}
  e_a = \pl_a = e_a\^m \pl_m
\end{equation}
for tangent vectors. Many geometric quantities take a particular
simple form involving the anholonomicity coefficients $c_{ab}\^c$
defined by the commutator of the basis of vector fields
\begin{equation}
  \left[ e_a, e_b \right] = c_{ab}\^c e_c.
\end{equation}
From (\ref{eanhol}) they are expressed by the inverse zweibein
\begin{equation}
  \label{eanho}
  \begin{aligned}
    c_{ab}\^c &= \left( e_a\^m \pl_m e_b\^n - e_b\^m \pl_m e_a\^n
    \right) e_n\^c \\
    &= -e_a\^m e_b\^n \left( \pl_m e_n\^c - \pl_n e_m\^c \right).
  \end{aligned}
\end{equation}
In two dimensions the anholonomicity coefficients are in one-to-one
correspondence with their own trace
\begin{equation}
  \label{eanhotr}
  c_b = c_{ab}\^a, \quad c_{ab}\^c = \dl_a\^c c_b - \dl_b\^c c_a.
\end{equation}
Eq.\ (\ref{eanho}) shows that the anholonomicity coefficients are
invariant under general coordinate transformations. Under the local
Lorentz boost with parameter $\om(x)$ they transform like a connection
\begin{equation}
  \dl c_b = -\om \e_b\^c c_c - \e_b\^c \pl_c \om.
\end{equation}
Some useful relations are
\begin{equation}
  \e^{ab} \pl_a e_b\^m = \e^{mb} c_b, \quad \e^{mn} \pl_m e_n\^a =
  -\e^{ab} c_b.
\end{equation}
Here the transformation from holonomic to anholonomic indices is
performed by using the two-dimensional zweibein (not the
supervielbein).

Apart from the zweibein a two-dimensional space-time is characterized
by the Lorentz connection which may be written in anholonomic
coordinates
\begin{equation}
  \om_a = e_a\^m \om_m.
\end{equation}
Then the two-dimensional curvature tensor,
\begin{equation}
  \widehat R_{mna}\^b = (\pl_m \om_n - \pl_n \om_m) \e_a\^b,
\end{equation}
yields the scalar curvature
\begin{equation}
  \label{escur}
  \widehat R = 2 \e^{ab} \pl_a \om_b + 2 \e^{ab} c_a \om_b.
\end{equation}
In two dimensions the trace of the torsion tensor $\tor_b =
\tor_{ab}\^a$ determines the full torsion tensor
\begin{equation}
  \tor_{ab}\^c = \dl_a\^c \tor_b - \dl_b\^c \tor_a
\end{equation}
and in the anholonomic basis becomes
\begin{equation}
  \tor_b = -c_b + \e_b\^c \om_c.
\end{equation}

Both zweibein and Lorentz connection are independent variables.  For a
given zweibein one can always construct a second Lorentz connection
and other geometrical quantities corresponding to zero torsion. The
latter condition makes the Lorentz connection depend on the zweibein
and its first derivatives through the coefficients of anholonomicity
(\ref{eanho})--(\ref{eanhotr})
\begin{equation}
  \tilde\om_a = \e_a\^b c_b.
\end{equation}
Here and everywhere else the tilde sign means that the corresponding
geometric quantity is derived for zero torsion.  The difference
between the two connections is given by the torsion tensor
\begin{equation}
  \label{erelco}
  \om_a = \tilde\om_a + \e_a\^b \tor_b.
\end{equation}
From (\ref{escur}) the scalar curvature corresponding to zero torsion
can be expressed through the anholonomicity coefficients as well:
\begin{equation}
  \widetilde R = 2 \pl_a c^a + 2 c_a c^a
\end{equation}

We use covariant derivatives $\widehat\nb_a = e_a\^m \widehat\nb_m =
e_a\^m (\pl_m + \om_m)$ and $\widetilde\nb_a = e_a\^m \widetilde\nb_m
= e_a\^m (\pl_m + \tilde\om_m)$ for both types of Lorentz connections
denoting the torsionless case by the tilde sign.  They are simply
related due to equation (\ref{erelco}) and used alternatively
according to the environment.

A sample relation exists between the two scalar curvatures:
\begin{equation}
  \widehat R = \widetilde R + 2 \widetilde\nb_a \tor^a = \widetilde R
  + 2 \widehat\nb_a \tor^a - 2 \tor_a \tor^a
\end{equation}


\section*{Acknowledgement}

We thank Peter van Nieuwenhuizen for discussions. This work has been
supported by Fonds zur F\"orderung der wissenschaftlichen For\-schung
(FWF) Project No.\ P 10221-PHY.  One of the authors (M.~K.) thanks the
Erwin Schr\"odinger International Institute where this work began, the
Austrian Academy of Sciences and the Russian Fund of Fundamental
Investigations, Grant RFFI-96-010-0312 and RFFI-96-15-96131, for
financial support.



\end{appendix}

\end{document}